\documentclass[preprint,review,12pt]{elsarticle}

\usepackage{graphicx}
\usepackage{amsmath}
\usepackage{amssymb}
\usepackage{hyperref}
\usepackage{float}

\biboptions{comma,square,numbers}

\journal{}

\begin{document}

\begin{frontmatter}

\title{Uncritical polarized groups: The impact of spreading fake news as fact in social networks}


\author[upm]{Jes\'us San Mart{\'\i}n}
\ead{jesus.sanmartin@upm.es}
\author[uniovi]{F\'atima Drubi}
\author[uned]{Daniel Rodr{\'\i}guez P{\'e}rez}

\address[upm]{Dept. Matem\'aticas del \'Area Industrial, ETSIDI, Universidad Polit\'ecnica de Madrid, Madrid, Spain}
\address[uniovi]{Dept. Matem\'aticas, Facultad de Ciencias, Universidad de Oviedo, Oviedo, Spain}
\address[uned]{Dept. F{\'\i}sica Matem\'atica y de Fluidos, Facultad de Ciencias, Universidad Nacional de Educaci\'on a Distancia, Madrid, Spain}

\begin{abstract}
The spread of ideas in online social networks is a crucial phenomenon to understand nowadays the proliferation of fake news and their impact in democracies. This makes necessary to use models that mimic the circulation of rumors. The law of large numbers as well as the probability distribution of contact groups allow us to construct a model with a minimum number of hypotheses. Moreover, we can analyze with this model the presence of very polarized groups of individuals (humans or bots) who spread a rumor as soon as they know about it. Given only the initial number of individuals who know any news, in a population connected by an instant messaging application, we first deduce from our model a simple function of time to study the rumor propagation. We then prove that the polarized groups can be detected and quantified from empirical data. Finally, we also predict the time required by any rumor to reach a fixed percentage of the population.
\end{abstract}

\begin{keyword}
online social network \sep fake news \sep rumor propagation \sep uncritical senders group
\end{keyword}

\end{frontmatter}


\section{Introduction}
The extraordinary global increase in online social network usage, and the ease of sending messages ubiquitously and almost instantaneously, have provided a fertile ground for the dissemination of fake news \cite{FigOli2017,MetaxasMustafaraj2010}. In addition, the use of these platforms can inevitably have deep social and political consequences. For instance, a disinformation campaign became a topic of wide public concern during the Brexit referendum in the UK, as well as during the US presidential election, both in 2016 \cite{AllGen2017, FigOli2017,SilSin2016}.

The seriousness of the situation due to the current and growing phenomenon of fake news lies not only in the speed and ease of reaching broad and very different strata of society but also in the negligible cost of this new form of worldwide interactive communication. This is a novel and simultaneously complicated situation that makes it necessary to study rumor propagation models in order to design appropriate countermeasures and avoid, for example, their potential impact on destabilization of liberal democracies \cite{MetaxasMustafaraj2010}.

Nevertheless, online social networks cannot be considered solely as mere media responsible for the propagation of fake news, since first they work in an inverse manner spreading real news. For this reason, understanding how ideas spread in social networks, as an element of both economic and political marketing, becomes a fundamental task. Companies, as well as political parties, that are interested in promoting their products or disseminating their ideas to the population, can benefit from the use of new and cheaper communication channels based on mobile applications such as WhatsApp (with millions of users worldwide), which are much more profitable than traditional marketing strategies, publicity campaigns, etc.

In order to avoid the destabilizing effects due to the spread of fake news, it is necessary to provide a suitable propagation model of rumors. In this sense, it is essential to first identify the fundamental variables that characterize this phenomenon. Given that there are great similarities between the spread of news and rumors and the spread of infectious diseases, a classic approach to study how information is disseminated is to define epidemiological models of populations \cite{Tambuscioetal2015,Jinetal2013}. Additionally, other studies focused on identifying the most efficient spreaders in a network \cite{Kitsatetal2010}. In \cite{BorgeHolthoeferetal2012}, two models were presented with the assumption that either spreaders are not always active or an ignorant is not interested in spreading the rumor. The effect of homogeneity and polarization, i.e. echo chambers, in the spreading of misinformation online was studied in \cite{DelVicarioetal2016}. Finally, why rumors spread so quickly in social networks was analyzed in \cite{Doerretal2012}. It should be noted that an element common to all these different approaches is that it is necessary to assume certain more or less believable hypotheses.

In this study, we present a news dissemination model base on a probabilistic approach that allow us to use the law of large numbers to determine the probability function of sending messages. To the best of our knowledge, there is no similar result for rumor spread models and we can only refer to the discussion in \cite{DaleyKendall1965}. One of our main contributions is that additional unrealistic hypotheses are not necessary. Therefore, we provide a more general rumor propagation model, which is also robust, without any limitation due to simplifications in the hypotheses. 

In addition to reducing the number of necessary hypotheses, another major issue when modelling is to obtain a faithful description of the reality that, in the particular case of social networks, is manifested by its heterogeneity. In our model, this heterogeneity is reflected in the fact that people are related through groups, with different sizes and not with the same sensitivity to the propagation of a message. In particular, the strongly polarized like-minded population groups are one of the key elements to amplify the spread of rumors. People belonging to this type of group will forwa all kind of news (false or not) with the sole criterion that they are related to their own ideological line.

It is interesting to note that news dissemination models hardly take into account the different reactions that people show when faced with true and false news. In this sense, there are two underlying problems for any model: the first refers to the different reaction of people to the veracity/falsity of the news and the second is related to how the certainty of the news is validated. In order to give an answer to these problems with our model, we simply focus on the propagation of the message. Thus, the authenticity or falsity of the message will be determined solely by the probability of propagation. Note that, except in very specific cases such us ``the president has been killed'', the information is judged according to the sender and is considered true based only on the idiosyncrasy of the person who receives it (similarly to those people who belong to the same ideological line will do it). On the contrary, that information will be considered false by those who have a different ideology, opposed to the sender's. The continuous variations of the ideological spectrum in the population imply the need to constantly evaluate the truth/falsity of a message. This phenomenon can be perfectly described by means of the different probabilities of propagation of the message: the more inclined the population is towards the conservatives, the greater the probability of spreading news that damages the image of the liberals and vice versa.

In fact, the probability of news propagation is a fundamental parameter when modeling the dissemination of ideas, since, if most of the individuals in a network propagate a message with low probability but a group, which we call the group of uncritical senders (USG), does so uncritically with probability $100\%$, the USG will be identified as a very polarized group.

The USG label is surely well defined in the sense that humans, not robots, accelerate the spread of false news more than the truth \cite{VosRoyAra2017}. The fact that people react differently when they receive the news, mainly depending on whether or not they have an ideological affinity, makes the probability of news propagation a key parameter to present and analyze our results. 

We must be aware that social networks are dynamic, that is, they are constantly changing. Because of this, the proposed models can only be useful if they can show how the spread of news changes when the relevant parameters of those models vary continuously. That information must be provided, as we will, with an analytic formulation. Therefore, we present a predictable model that can be used to eliminate or mitigate the dangerous consequences of spreading fake news as, for instance, biasing the vote in government elections, misallocating resources after natural disasters or terrorist actions, misguiding in the investment measures after the stock market crash, etc. All this with a great political, social and economic impact, due to the fact that the number of people who are currently only informed of the news through their social networks is increasing.

Finally, given that it is a well-known fact that news does not necessarily have to be disseminated from a single source, but that there may be different sources from which the same news spreads in cascade to the entire social network, we work with our propagation model of rumors but using different initial conditions, that is, different seeds are considered in the population. These seeds represent different groups of individuals that initiate the propagation of the same rumor in the network.

Our results show how the distributions of the propagation probability of news in social networks change over time as a function of polarized groups of uncritical senders. Particularly, we found that the probability of spreading rumors varies from an exponential evolution to a logistic one. As a result, simply observing how fake news is spread in a social network, we can detect with our model the presence of a polarized group of uncritical individuals, which becomes a very useful tool to design countermeasures to deactivate such groups, if necessary.

The rest of this paper is structured as follows. In section \ref{Sect_Methodology}, we present a model of rumor propagation in a social network based on WhatsApp. This is a numerical model defined from empirical data. In particular, the standard distributions of the number of person-to-person contacts and of group sizes were estimated experimentally. In addition, the key parameters that allows us to analyze the phenomena of spreading rumors are identified and their values are also estimated. In section \ref {Section_Numerical_Results}, analytic expressions that fit the numerical simulations are deduced. Based on these expressions, a theoretical model is proposed that captures the observed behavior and sets the basis to interpret the dynamics that led to those results. In section \ref{Sect_Discussion}, we discuss the potential applications of our model and its predictive capabilities. We conclude the paper in section \ref{Sect_Conclusions}.

\section{Methodology}
\label{Sect_Methodology}

Taking into account that WhatsApp is one of the most popular messaging application all over the world (e.g. it reaches a $70\%$ of the total population in Spain \footnote{https://www.messengerpeople.com/global-messenger-usage-statistics/\#Spain}), we simulate our rumor propagation model in a social network based on this application. In order to properly simulate message spreading over this network, we must distinguish two types of contacts, person-to-person and groups. We also need to know the standard distributions of the number of person-to-person contacts and of group sizes. Both were estimated experimentally as described below.

\subsection{\label{subsec:whatsapp-dist}Statistical characterization of a WhatsApp network}

A sample of 150 college students (age range 18-20 years) was used to obtain the statistical distribution of person-to-person and groups in the WhatsApp network. Individuals were asked about the number of contacts in their cell phone contact lists, the number of WhatsApp groups with three or more members and the sizes of those groups (number of members including themselves). Moreover, they were asked about the number of frequent individual WhatsApp contacts, defined as those to whom they would text with a controversial message they learned from any source.

In any contact list, there are many contacts that do not belong to any group (meaning by group a set of persons linked by a common interest). Hence, we posed the following question: in case they learned about a particularly ``hot news'', to whom of their contacts would they expressly text for commenting on it? Obviously, that ``hot news'' would not be sent to their attorney or plumber and, depending on the topic, neither to a relative. Thus, we define the groups of size two as those contacts in the contact list that, although they may be members of any WhatsApp group, would receive a message directly from the user about a news considered especially relevant. Only these  size-two groups constitute the set of person-to-person links of each WhatsApp user.

From our sample data, we find out that the number $m$ of person-to-person contacts an individual has follows approximately a normal distribution
\begin{equation}
    \label{Dist_Normal}
    N(m,\mu, \sigma) = \frac{1}{2\pi\sigma} e^{-\frac{(m-\mu)^2}{2\sigma^2}}
\end{equation}
where $\mu=7.35$ and $\sigma=4.38$. 

\begin{figure}
    \centering
    \includegraphics[width=0.48\textwidth]{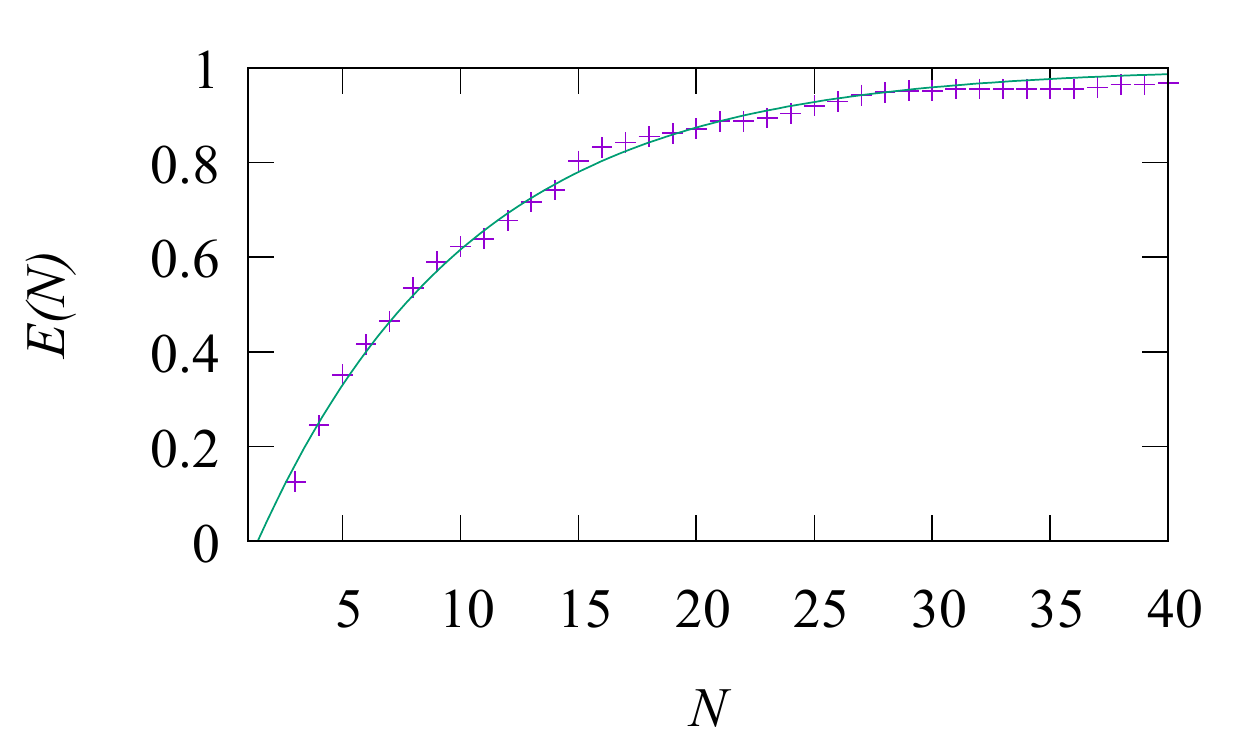}
    \includegraphics[width=0.48\textwidth]{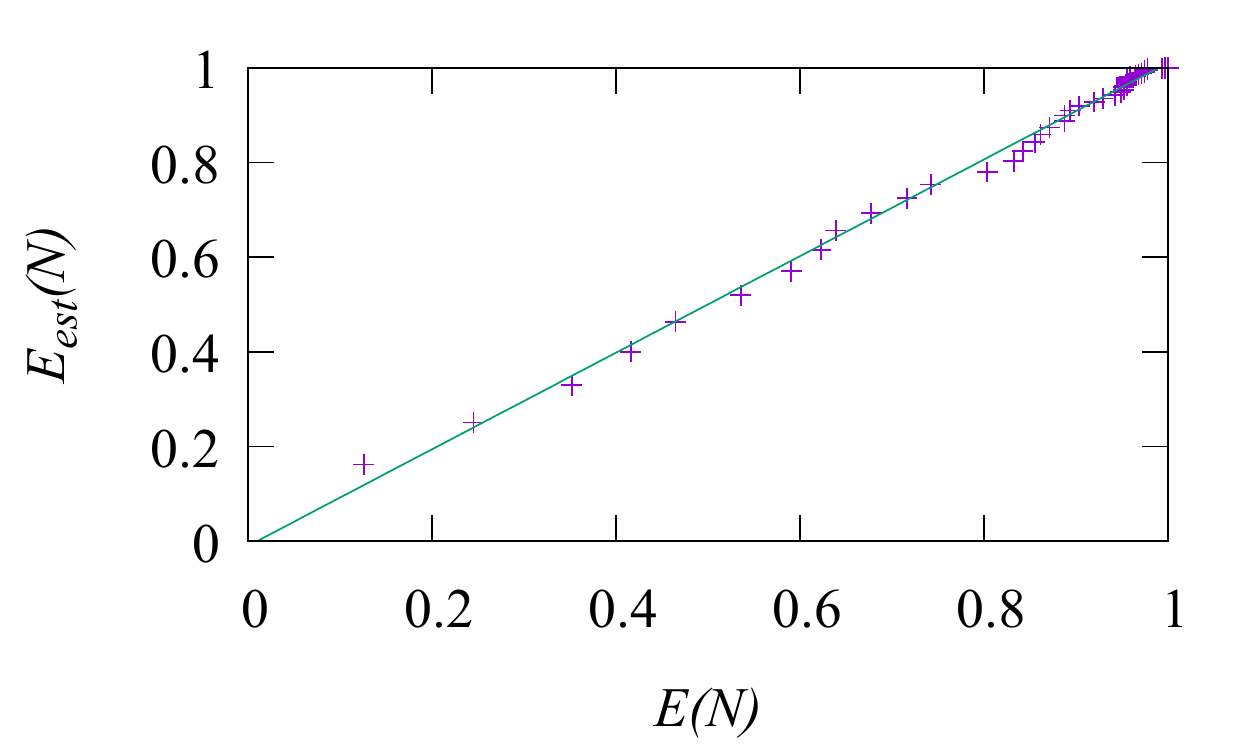}
    \caption{\label{fig:empirical_sample} WhatsApp empirical group size distribution (left) and the goodness of fit of the cumulative distribution function (right) to an exponential ($r^2 = 0.9959$). $N$ represents the number of individuals in a group ($3\leq N\leq 30$), $E(N)$ represents the empirical distribution function, i.e. the fraction of groups of size $\leq N$, and $E_{est}(N)$ is the estimation of $E(N)$.}
\end{figure}

The group size distribution, for sizes of $3 \leq N \leq 30$ members (that included the majority of our samples), fits an exponential distribution
\begin{equation}
\label{Dist_Exp}
E(N)=1-\exp(-\lambda (N-a))
\end{equation}
where coefficients are adjusted from the data as $\lambda = 0.1113 \pm 0.0019$ and $a = 1.41 \pm 0.12$ ($r^2 = 0.9959$). See Figure \ref{fig:empirical_sample}.

In summary, the groups between 3 and 30 individuals in size are characterized by an exponential distribution function, while the groups of size 2, i.e. person-to-person links, are characterized by a normal distribution function.

\subsection{The model}

In this paper, we introduce a novel model where individuals are linked either by person-to-person relations or by belonging to the same WhatsApp group.
These links fit the distributions obtained in section \ref{subsec:whatsapp-dist}. It is worth noticing that, initially, a given fraction of population, named the seed,  knows a rumor. This rumor may spread to other linked individuals at each iteration during the numerical simulations. The rumor propagation proceeds iteratively taking into account two main rules. First, the individual propagation probability at each iteration is given by a uniform distribution. Second, there may also be a group of individuals, the so-called uncritical senders group, which always propagates the rumor at each iteration, i.e. with an individual propagation probability of $100\%$. 

As the numerical simulations were repeated independently a large number of times and then we averaged the result, the law of large numbers guarantees that the final law of distribution followed by the rumor propagation is determined as well as the time required for a rumor to reach a given fraction of the population. An important feature of the model is that it shows when a population is polarized regarding a given rumor. 

For the sake of enabling a better understanding of the model that is developed in detail below, we provide here some useful definitions:
\begin{description}

\item{Burned:} A person who knows the rumor is said to be a burned individual. Observe that if an individual is burned, he remains in that state until the end of the rumor propagation process. 

\item{Sender:} A person who knows the message (a burned individual) and texts it to any of his contacts is called a sender.

\item{Receiver:} Any individual (burned or not) in the population that is reached by the message is said to be a receiver.

\item{Seed:} All those individuals who know the rumor at the beginning of its propagation constitute the seed, i.e. the set of burned individuals at the initial time $t=0$.

\item{Person-to-person relation:} A direct link between two individuals of the population is called a person-to-person relation. This relationship is singled out from the group relationship because it represents a one-way link. If an individual $i$ is connected with another individual $j$, then $j$ may receive text messages sent by $i$, but this does not imply that individual $j$ could also send messages to $i$.

\item{Group:} The set of three or more individuals in the population that are interconnected, meaning that what one of the individuals sends to the group is received by all others in the group simultaneously.

\item{Individual propagation probability ($P_{IP}$):} The probability of the rumor being sent by a burned individual, who knows the rumor, to one of its contacts (person-to-person probability) or groups in the WhatsApp network.

\item{Individual initial probability ($P_{II}$):} The probability that individuals are part of the seed. That is, the probability that an individual knows the message at the beginning of the process of spreading the rumor.

\item{Uncritical senders group (USG): } The set of individuals who automatically send the message to all of their contacts when they receive it is named the uncritical senders group. Hence, $P_{IP}$ is always equal to $100\%$ for individuals in the USG. 

\item{USG membership probability ($P_{USG}$): } The probability that individuals belong to the USG. It represents the fraction of the population that forms the USG. The size of the USG is fixed as an initial condition and once fixed, and randomly chosen the individuals that are a part of it, it does not change.

\end{description}

\subsection{\label{subsection:Algorithms}Model algorithms} 
The model consists of two main steps. First, an algorithm that generates populations of connected individuals as described in section \ref{subsec:whatsapp-dist}. Second, another algorithm simulating the spreading rumors among these populations.

\subsubsection{Populations}
In order to simulate a network with the statistical properties described in section \ref{subsec:whatsapp-dist}, we proceeded to implement an algorithm that reproduces that structure.  That is, given a population of $Np$ individuals, we establish connections between pairs of individuals following the normal distribution (\ref{Dist_Normal}) and also generate groups of 3 to 30 interconnected individuals with group sizes given by (\ref{Dist_Exp}).

When come to consider group construction, we will separate groups formed by just 2 individuals from groups formed by 3 or more individuals; the reason is that in two-person groups, connections are not bidirectional (as we explained above), whereas for groups with 3 or more individuals it is.

We will assume a normal distribution (\ref{Dist_Normal}) for the number of 2-groups an individual belongs to, having the mean and standard deviation fitted from our sample.
From that distribution, the number of individuals $n_m$ that are expected to have a number $m$ of person-to-person connections was computed, for $m=0,1,\dots,30$. That is, first the number of individuals in the population having zero person-to-person contacts, $n_0=N(0,\mu,\sigma)$ was computed and then, recursively, the number of individuals having $m$ person-to-person contacts $n_m=N(m,\mu,\sigma)-N(m-1,\mu,\sigma)$ for $m=1,\dots,30$.
Then the $n_m$ individuals having $m$ contacts were chosen randomly from the population, linking them to other individuals also randomly chosen. Notice that these relationships, by their very nature, are not bidirectional given that they are not proper WhatsApp groups, but the initial individual will send messages to the destination individual, but that does not have to work in the reciprocal.

To compute the number of groups of size 3 or more an individual belongs to, we will use the exponential distribution in (\ref{Dist_Exp}) with the parameters fitted from our sample. Specifically, the number of groups of size $i$ is calculated as $E(i)-E(i-1)$ for $i=3,4,\dots,N$. Once known the number of groups of each size (from 3 to 30 individuals), the individuals of the population belonging to each group are randomly chosen.

Thirty populations with 10000 individuals each were simulated following the previous procedure, taking into account that only 70\% of the individuals (the penetration of WhatsApp in Spanish population) should be connected either person-to-person or to groups of sizes 3 to 30. Each of these simulations provided a network of connections between individuals of the population.

\subsubsection{Message spreading}

Once the population has been simulated (see above) the rumor spreads among its individuals according to the following rules:
\begin{enumerate}
    \item The fraction of the population that knows the rumor, that is the initial seed, is chosen initially from the connected population given by $P_{IP}$.
    \item One individual belonging to the see is randomly chosen 
    \item The groups to which that individual belongs are identified
    \item That individual will pass the rumor to each of these groups with a probability $P_{IP}$; for this, a uniformly distributed random value in $[0,1)$ is generated for each group, and the rumor is spread to the group if that value is less than $P_{IP}$
\end{enumerate}
Steps 2 and 3 are repeated until all seed individuals are exhausted.
\begin{enumerate}
    \item[5. ] All the individuals that know the rumor become part of the seed, and the propagation process is repeated 100 times; this number of iterations is chosen because in most simulations the rumor reaches the entire population before that time.
\end{enumerate}

Once the initial seed and the USG have been chosen, according to the individual initial probability ($P_{II}$), the uncritical senders group membership probability ($P_{USG}$), and the individual propagation probability ($P_{IP}$) that characterize the initiation and propagation of the messages on the social network, the algorithm proceeds according to the following rules:
\begin{enumerate}
    \item[i)] The rumor is propagated among the individuals in the population for 100  iterations (the simulation unit of time).
    \item[ii)] At each iteration, every burned individual in the population, that is, every individual who knows the rumor, will communicate it to each of its contacts with  probability $P_{IP}$.
    \item[iii)] If an individual belongs to the USG and is burned, it will propagate the message to all of its contacts (that is, $P_{IP}$ equals 1 for individuals in the $USG$). 
    \item[iv)] After every iteration, burned individuals  (and WhatsApp groups) are recorded and counted.
\end{enumerate}

Given that message transmission is a random process, message spreading algorithm is run 50 times on each population using, for that population always the same initial seed. The result after finishing each run is the series of numbers of burned individuals at each iteration. The final result is summarized as the average number of these 50 series. Along the same lines, given that each population is computed as a random sample from a probability distribution, in section \ref{Section_Numerical_Results} we will work with the results obtained by averaging over the 30 populations, in order to characterize the temporal evolution of the average number of burned individuals for each set of parameters $P_{IP}$, $P_{II}$ and $P_{USG}$.

\section{Numerical results} \label{Section_Numerical_Results}
To proceed with the simulations, the parameter space $P_{II}\times P_{IP} \times P_{USG}$ is sampled at $500$ points given by the following tuples $(u,v,w)$ where
$u, v\in \{k\cdot 10^{-2}: k = 1,2,\dots,10\}$ and $w\in \{k\cdot 10^{-2}:k=0,3,5,7,10\}$.
Each of these $500$ points characterizes individuals and groups in a simulated population. For each of the $500$ points, $30$ such populations with $10000$ individuals each were randomly generated following the algorithm described in \ref{subsection:Algorithms}. Notice that although these populations are different, they are statistically the same because they were generated using the same parameters at the same point of the parameter space. For each of these $30$ populations, a news was spread over the network during $100$ iterations, and this random process was repeatedly simulated $50$ times, choosing a different initial condition for each of the $30$ populations (the seed individuals who know the information at the beginning of the propagation). The propagation time series was computed as the average of the $50$ simulations for a given population, resulting in $30$ averaged spreading evolutions corresponding to the $30$ statistically identical populations. Finally, the average over the $30$ populations was computed as the propagation time series corresponding to one of the $500$ points in the parameter space.

The goal of this numeric modeling is to fit an analytic expression that summarizes the spreading of the news corresponding to the model described above.

\subsection{Case $P_{USG}=0$}

To see the effect polarized groups have on news spreading, it was first needed to know how news spread without those groups ($P_{USG}=0$). In this section these results are shown and an analytic expression fitted to them.

The analysis is based on data from simulations obtained as the average number of burned individuals at the $n$-th iteration taking into account all populations (30) and all 50 simulations performed on each population. That is, the behavior of the function corresponding to the average number of burned individuals at each iteration is studied:
$$f(n) = \frac{\mbox{Total of burned individuals at n-th iteration} }{ \mbox{no. of populations} \times \mbox{no. of simulations} \times N_p}$$
where $N_p$ is the number of individuals of the population connected by WhatsApp, and $n = 1, \dots, 100$.

The space of parameters is sliced in sections, that is, we study how rumor spread in populations for different values of $P_{II} \in [0, 0.1]$ and, for each of these values, the behavior for different values of $P_{IP} \in [0, 0.1]$ is also studied.  

\begin{figure}
    \centering
    \begin{tabular}{cc}
    {\tiny $P_{II}=3\%$} & {\tiny $P_{IP}=2\%$} \\
    \includegraphics[width=0.48\textwidth]{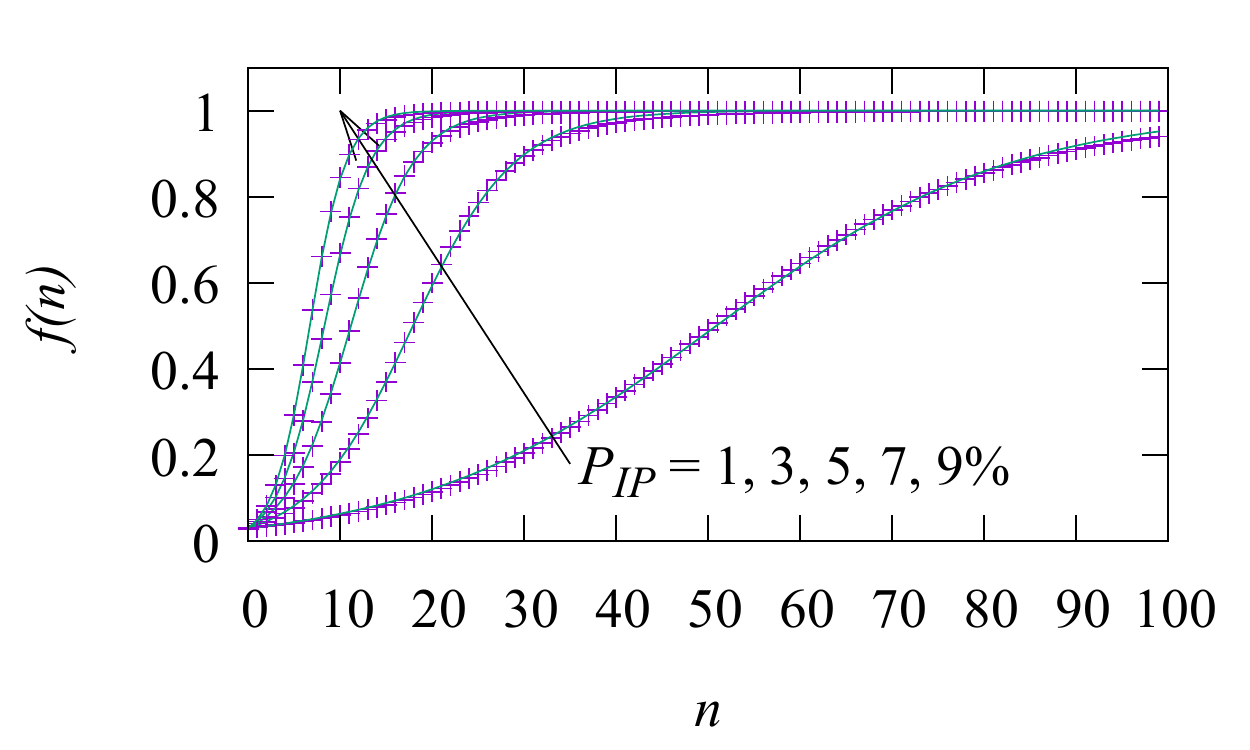} &
    \includegraphics[width=0.48\textwidth]{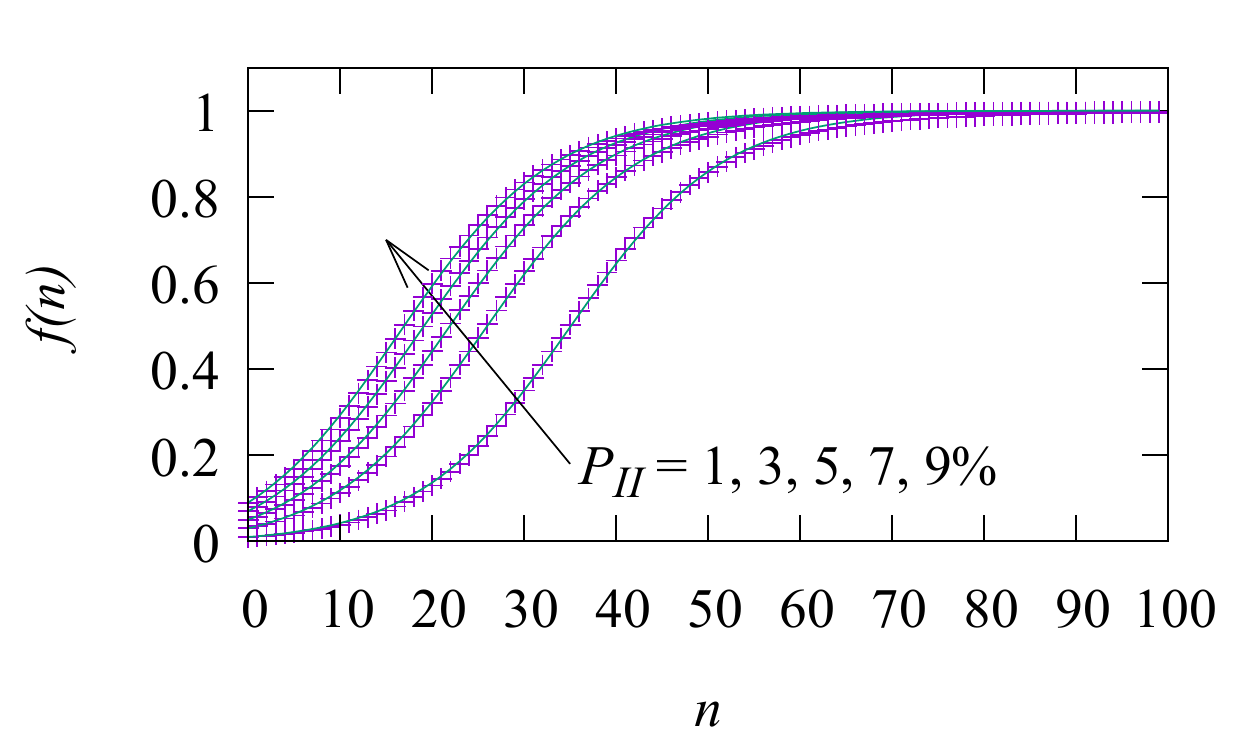}
    \end{tabular}
    \caption{\label{fig:ft_PPI_PII} Average number of burned individuals $f(n)$ at each iteration $n$ for different values of $P_{IP}$ (left, $P_{II}=3\%$) and $P_{II}$ (right, $P_{IP}=2\%$). Fixed parameter: $P_{USG}=0$. The blue lines correspond to the fitting expression (\ref{eq:fit_num_Fn}) in each case.}
\end{figure}

The simulation results $f(n)$ (see Figure \ref{fig:ft_PPI_PII}) picture the average evolution of the fraction of burned individuals after successive iterations (indexed by $n$) that approach a continuous function $F(t)$ dependent of time $t$.

For values of $P_II$ around $10\%$, function $F(t)$ is well approximated by an exponential of the form
$$ F_{exp}(t) = 1 - A \exp(-t/a)$$
whereas for smaller values of $P_{II}$ a better fit is obtained using a logistic function
$$ F_{log}(t) = \frac{1}{1 + B \exp(-2t/a)} $$
Therefore, we construct a function that fits the data in the entire range by changing from one form to the other through a parameter $\epsilon$:
\begin{equation}\label{eq:fit_num_Fn}
    F(t)=\frac{ C \exp[(1+\epsilon) (t-b)/a] - 1 }{ \frac{2 \epsilon }{1-\epsilon }+C \exp[(1+\epsilon) (t-b)/a] }
\end{equation}
where $a$, $b$, $\epsilon$ and $C$ are the fitted coefficients whose values depend on simulation parameters $P_{II}$ and $P_{IP}$ (how, will be shown in what follows). Coefficient $a$ represents a characteristic time scale. Coefficient $b$ has the meaning of a time origin, thus can be taken as zero.

To fulfill the initial condition $F(t=0)=P_{II}$ (the initial fraction is just the seed), $C$ had to be taken as a function of $a$, $b$, $\epsilon$ and $P_{II}$ given by:
\begin{equation}\label{eq:fit_num_C}
    C(a,b,\epsilon)= \frac{ 1 + \frac{2 \epsilon }{1-\epsilon } P_{II} }{ ( 1 - P_{II} )  \exp[ (1+\epsilon) b/a ]}
\end{equation}

From the fitting of the simulation data for different $P_{II}$ and $P_{IP}$, an expression is given for $\epsilon$ in the form:
\begin{equation}\label{eq:fit_num_ep}
    \epsilon = 1 + \frac{aa P_{II}}{1+\exp(bb/a)}
\end{equation}
where $aa$ is less than zero, in order to produce an $\epsilon$ between 0 and 1.
Likewise, $a$ is fitted to:
\begin{equation}\label{eq:fit_num_1prea}
   1/a = cc \, P_{IP}^{ee}
\end{equation}
which leads to a characteristic time $a$ tending to infinity, when the probability $P_{IP}$ tends to zero (that is, if the propagation probability is very small, the time a rumour will take to spread over the entire network will become very large). The values of $aa$, $bb$, $cc$ and $ee$ can be found in table  \ref{tab:Coeficientes} (for $P_{USG}=0$).

The power law (\ref{eq:fit_num_1prea}) did not provide a good collapse of all the points towards the fitting curve (see Fig. \ref{fig:1a_PPIa}a). Thus, in order to achieve that collapse (see Fig. \ref{fig:1a_PPIa}b), a factor $(1+gg P_{II})$ is considered to correct for the small dependency that $a$ showed on $P_{II}$, resulting in an expression of the form
\begin{equation}\label{eq:fit_num_1a}
   1/a = cc \, P_{IP}^{ee} \, (1+gg \, P_{II})
\end{equation}

Inserting into (\ref{eq:fit_num_Fn}) the expressions for $C$, $\epsilon$ and $a$  given by (\ref{eq:fit_num_C}), (\ref{eq:fit_num_ep}) and (\ref{eq:fit_num_1a}) an expression for $F(t)$ is obtained. 

It is remarkable that all the resulting expressions depend exclusively of $P_{IP}$, $P_{II}$ and $P_{USG}$ so that the number of burned individuals at a given iteration $n$, are given by (\ref{eq:modelo_teorico}) in terms of just these parameters. Hence, the number of burned individuals can be computed at any time, given a point ($P_{USG},\,P_{IP},\,P_{II}$) in the parameter space, or vice versa, that point in the parameter space can be computed once the time required to burn a given fraction of the connected population has been fixed.

\subsection{Case $P_{USG} \neq 0$: Effect of the uncritical senders group}\label{sec:USGneq0}

\begin{figure}
    \centering
    \includegraphics[width=0.9\textwidth]{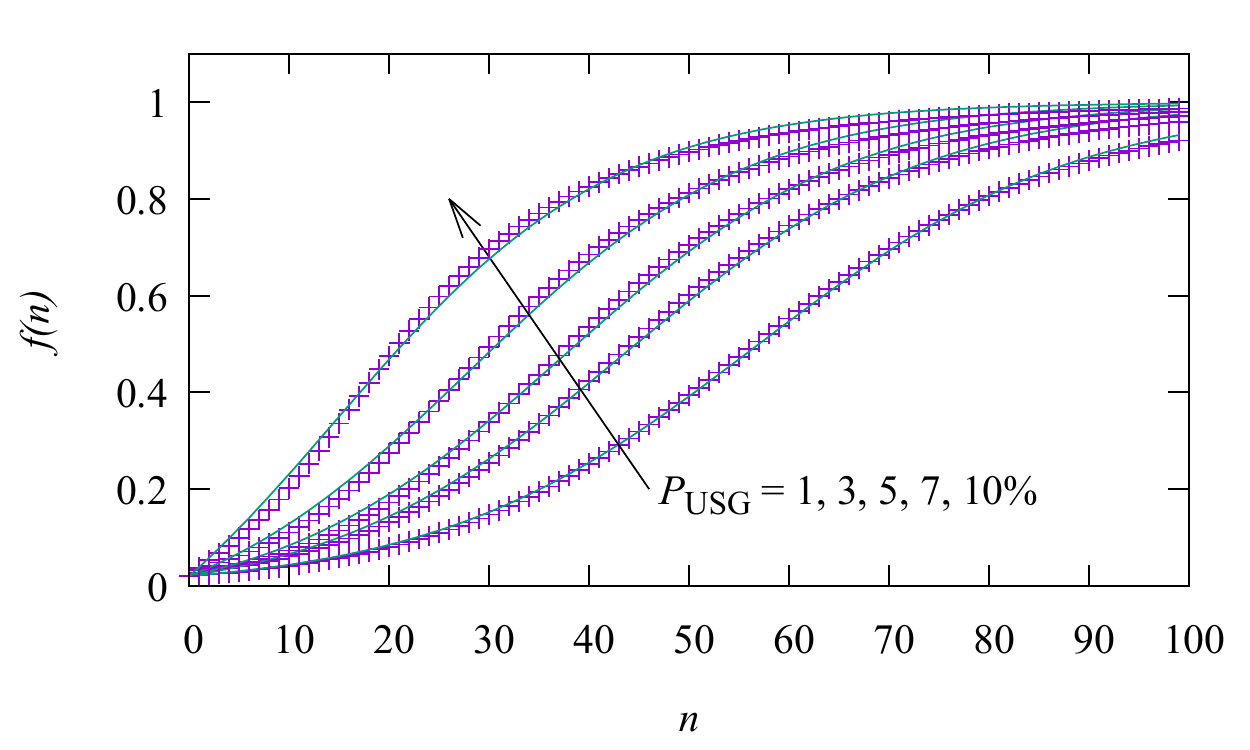}
    \caption{\label{fig:ft_USG} Average number of burned individuals $f(n)$ at each iteration $n$ for different values of $P_{USG}$. Fixed parameters: $P_{II}=2\%$ and $P_{IP}=1\%$. The results of fitting the simulation data to the analytic expression (\ref{eq:fit_num_Fn}) are also shown (blue lines).}
\end{figure}

Assuming the presence of very polarized groups among the population, what we call the uncritical senders group, it is necessary to see how their existence modifies the spread of news just described. The introduction of the USG affects the evolution of the fraction of the population a rumor reaches at a given time, as the simulations show. Some results of these numerical simulations are shown in Figure \ref{fig:ft_USG}, together with their respective fitting to expressions formally identical to (\ref{eq:fit_num_Fn}), which shows its validity also for $P_{USG} \neq 0$.

\begin{figure}
    \centering
    \includegraphics[width=0.9\textwidth]{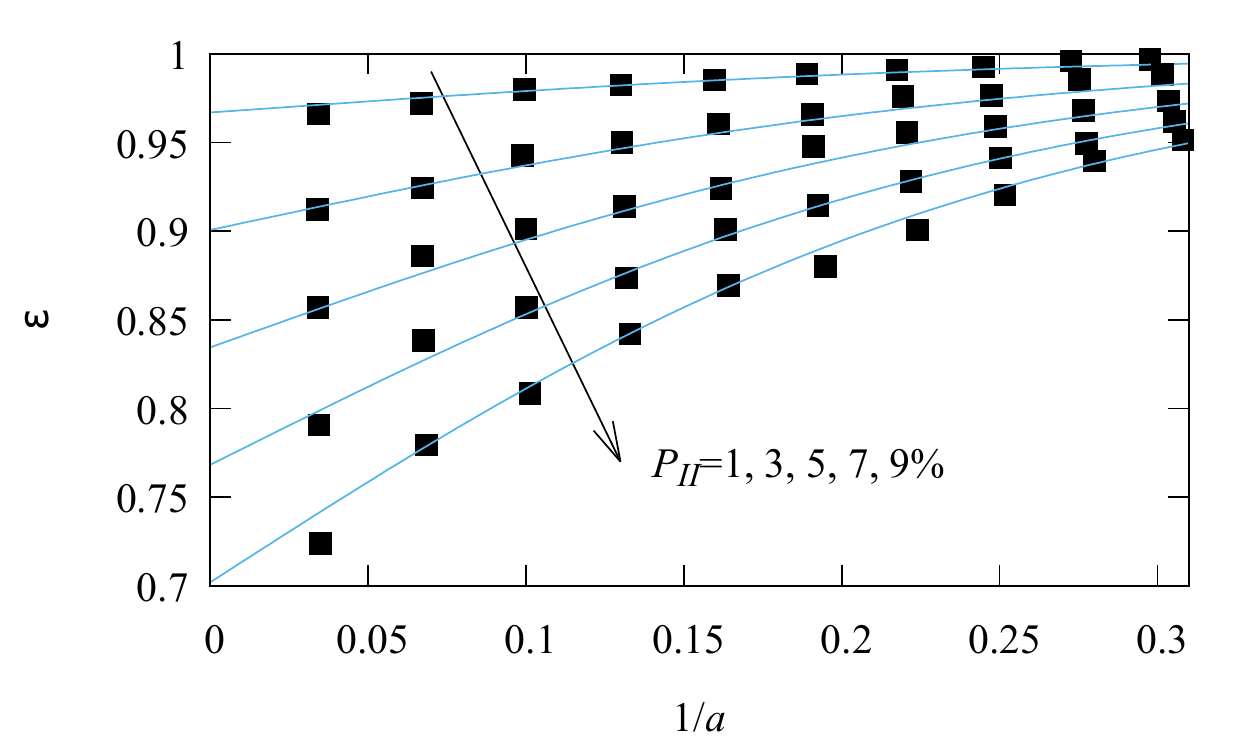}
    \caption{\label{fig:ep_1a} Fitted values of the coefficient $\varepsilon$ with respect to $1/a$ for different values of parameter $P_{II}$. Fixed parameter: $P_{USG}=3\%$. The graph of the fitted expression relating both coefficients with $P_{IP}$ is also shown (blue lines).}
\end{figure}

\begin{figure}
    \centering
    \begin{tabular}{ll}
    \includegraphics[width=0.47\textwidth]{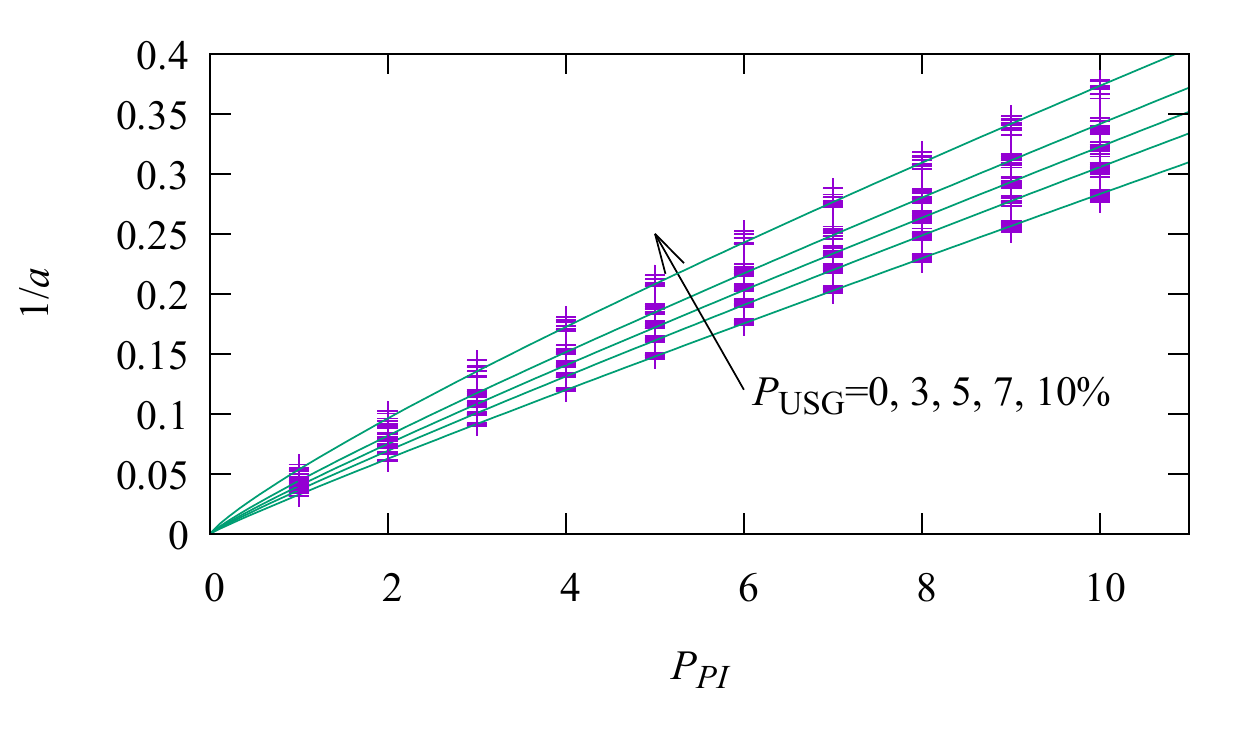} & \includegraphics[width=0.47\textwidth]{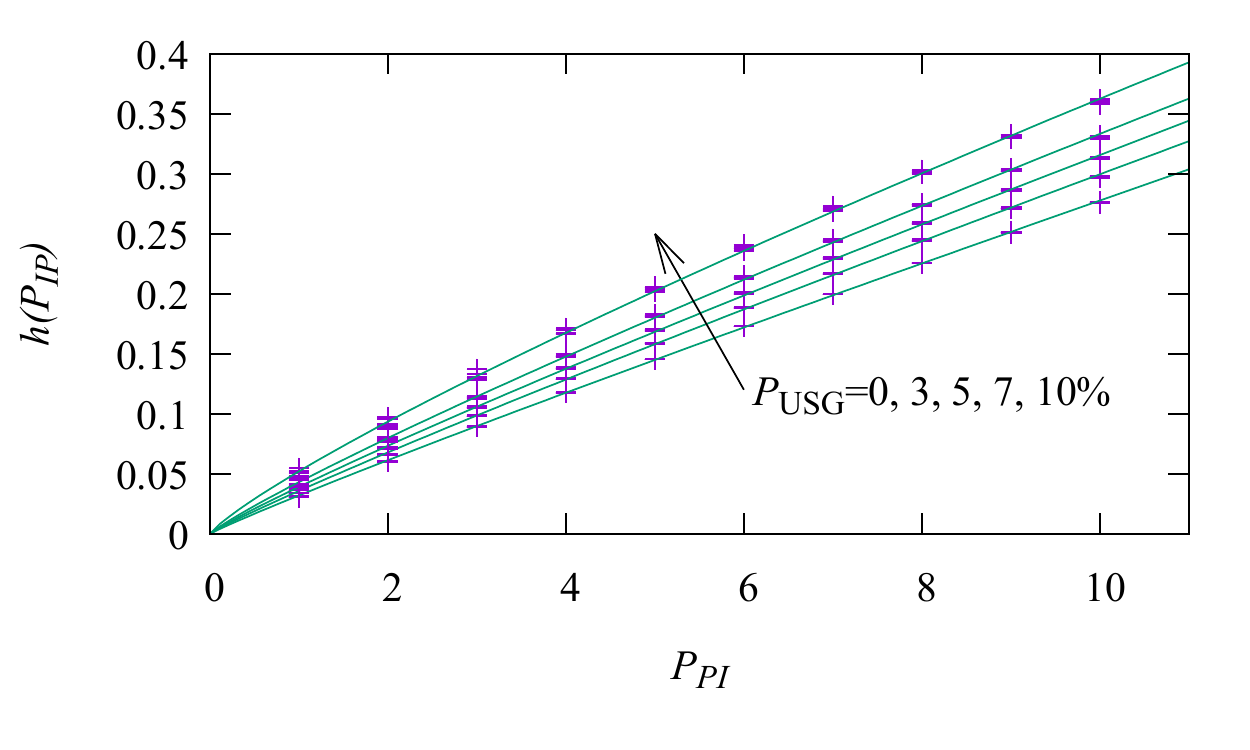}
    \end{tabular}
    \caption{\label{fig:1a_PPIa} Left: Fitted values of the coefficient $1/a$ for different values of $P_{IP}$ and $P_{USG}$ (and of $P_{II}$). The graph of the power law expression (\ref{eq:fit_num_1prea}) is also shown (blue lines). Right: Values of $h(P_{II})=\frac{1/a}{1+gg P_{II}}$ for different values of $P_{IP}$ and $P_{USG}$. Taking into account the effect of $P_{II}$ on $1/a$, the correction given by (\ref{eq:fit_num_1a}) is shown (green lines). Notice how the data now collapse to the modified power law.}
\end{figure}

The relationship (\ref{eq:fit_num_ep}) between $\epsilon$ and $1/a$ obtained above for $P_{USG}=0$ is still valid for $P_{USG} \neq 0$, according to the numerical results (see Figure \ref{fig:ep_1a} and table \ref{tab:Coeficientes}). Also expression (\ref{eq:fit_num_1a}) for $1/a$ in terms of $P_{IP}$ and $P_{II}$ remains valid (see Figure \ref{fig:1a_PPIa} and table \ref{tab:Coeficientes}).

\begin{table}[htb]
\centering
\begin{tabular}{c|c|c|c|c}
\cline{2-5}
     & \multicolumn{2}{c|}{aa}& \multicolumn{2}{|c}{bb} \\ \hline
    $P_{USG}$ & Values & Error & Values & Error  \\ \hline
    $0\%$ & $-0.0208 \pm 0.0003$ & 1.49\% & $7.32 \pm 0.16$ & 2.18\% \\
    $3\%$ & $-0.0663 \pm 0.0006$ & 0.91\% & $7.69 \pm 0.09$ & 1.23\% \\
    $5\%$ & $-0.1148 \pm 0.0010$ & 0.90\% & $8.28 \pm 0.09$ & 1.12\% \\
    $7\%$ & $-0.180 \pm  0.002$ & 1.12\% & $8.77 \pm 0.11$ & 1.31\% \\
    $10\%$ & $-0.328 \pm 0.006$ & 1.93\% & $9.47 \pm 0.19$ & 2.02\% \\ \hline \hline
     & \multicolumn{2}{c|}{cc} & \multicolumn{2}{|c}{ee} \\ \hline
    $P_{USG}$ & Values & Error & Values & Error\\ \hline
    $0\%$ & $0.03204 \pm 0.00011$ & 0.33\% & $0.9381 \pm 0.0015$ & 0.16\% \\
    $3\%$ & $0.03579 \pm 0.00015$ & 0.41\% & $0.9228 \pm 0.0019$ & 0.21\% \\
    $5\%$ & $0.0391 \pm 0.0002$ & 0.51\% & $0.907 \pm 0.002$ & 0.26\% \\
    $7\%$ & $0.0431 \pm 0.0002$ & 0.56\% & $0.888 \pm 0.003$ & 0.30\% \\
    $10\%$ & $0.0521 \pm 0.0004$ & 0.77\%  & $0.843 \pm 0.004$ & 0.43\% \\ \hline 
    \hline
     & \multicolumn{2}{c|}{gg} & \multicolumn{2}{c}{} \\ \cline{1-3}
    $P_{USG}$ & Values & Error & \multicolumn{2}{c}{} \\ \cline{1-3}
    $0\%$ & $0.00354 \pm 0.00019$ & 5.42\% & \multicolumn{2}{c}{} \\
    $3\%$ & $0.0036 \pm 0.0002$ & 6.70\% & \multicolumn{2}{c}{} \\
    $5\%$ & $0.0039 \pm 0.0003$ & 7.65\% & \multicolumn{2}{c}{} \\
    $7\%$ & $0.0045 \pm 0.0003$ & 7.54\% & \multicolumn{2}{c}{}\\
    $10\%$ & $0.0054 \pm 0.0005$ & 8.61\% & \multicolumn{2}{c}{}\\ \cline{1-3}
\end{tabular}    
    \caption{\label{tab:Coeficientes} Coefficients and exponents obtained from fitting expressions (\ref{eq:fit_num_ep}) and (\ref{eq:fit_num_1a}) to data from simulations with different values of $P_{USG}$.}
\end{table}

Nevertheless, the effect of considering a USG in the network makes that the coefficients $aa$, $bb$, $cc$, $ee$ and $gg$ in equations (\ref{eq:fit_num_ep}) and (\ref{eq:fit_num_1a}) become functions of $P_{USG}$. The functional expressions of $aa$, $bb$, $cc$, $ee$ and $gg$ were derived from the numerical results (for $P_{USG}\in[0, 0.1]$) as follows:
\begin{equation}\label{eq:fit_num_USG}
    \begin{split}
    aa &= aa_1 + aa_2 P_{USG} + aa_3 {P_{USG}}^2  \\
    bb &= bb_1 + bb_2 {P_{USG}}^4/(1+ bb_3 {P_{USG}}^3)    \\
    cc &= cc_1 + cc_2 P_{USG} + cc_3 {P_{USG}}^2 \\
    ee &= ee_1 + ee_2 P_{USG} + ee_3 {P_{USG}}^2 \\
    \end{split}
\end{equation}
where the values of the new coefficients are given in Table \ref{tab:fit_num_USG_coefs}. Substituting expressions (\ref{eq:fit_num_USG}) in equations (\ref{eq:fit_num_ep}) and (\ref{eq:fit_num_1a}), and these in equations (\ref{eq:fit_num_C}) and (\ref{eq:fit_num_Fn}), the general equation for the evolution of $F(t)$ is obtained, exclusively in terms of the parameters $P_{IP}$, $P_{II}$ and $P_{USG}$, as was our goal.

\begin{table}[htb]
    \centering

    \begin{tabular}{c|r|l}
    Parameter & Value & Error \\
    \hline{} 
    $aa_1$       &  -2.2  & $\pm$0.4 \\
    $aa_2$       &  -63  & $\pm$18   \\
    $aa_3$       &  -2410  & $\pm$170 \\
    \hline{} 
    $bb_1$      &   7.319  & $\pm$0.010  \\
    $bb_2$      &   1.09$\times10^6$  & $\pm$0.09$\times10^6$ \\
    $bb_3$      &   49000  & $\pm$4000  \\
    \hline{} 
    $cc_1$      &   2.405  & $\pm$0.010  \\
    $cc_2$      &   4.7  & $\pm$0.5  \\
    $cc_3$      &   -35  & $\pm$4 \\
    \hline{} 
    $ee_1$      &    0.9375  & $\pm$0.0017 \\
    $ee_2$      &   -0.25  & $\pm$0.08  \\
    $ee_3$      &   -7.0  & $\pm$0.7  \\
    \hline{} 
    $gg_1$      &  0.3541  & $\pm$0.0010 \\
    $gg_2$      &  10800  & $\pm$600 \\
    $gg_3$      & 4700   & $\pm$300 \\ \hline
    \end{tabular}

    \caption{ \label{tab:fit_num_USG_coefs} Coefficients of equations (\ref{eq:fit_num_USG}). }
\end{table}

\subsection{Dynamical spreading model}

In this section we present a theoretical model that captures the behaviors described above. For that, we will interpret our discrete time model in terms of a continuous time model that may be derived from a differential equation. For that, it is enough to see that $P_{IP}$ is a probability per unit time of a message being propagated by one individual to another or to its group, being that unit time the time separating one iteration from the next one. Expression (\ref{eq:fit_num_Fn}), describing the fraction of burned individuals as a function of time turns out to be a solution of the following differential equation:
\begin{equation}\label{eq:modelo_teorico}
dF/dt = \frac {2 \epsilon}{a} \left[ \frac{1-\epsilon}{2\epsilon}+  F \right] (1-F)
\end{equation}
If this equation is interpreted as a law of mass action, the coefficient $A=\frac {2 \epsilon}{a}$ would stand for the spreading velocity towards the unburned population  $(1-F)$ of news originating in the burned population $F$ plus an ``invisible'' population given by the additional term $G=\frac{1-\epsilon}{2\epsilon}$.

In order to find out the meaning of this  ``invisible'' population (which is constant along the spreading process), let us consider the case of $P_{II} \rightarrow 0$, for which $\epsilon \rightarrow 1$. Then, employing (\ref{eq:fit_num_ep}), equation (\ref{eq:modelo_teorico}) tends to
$$ dF/dt = 1/a \left[ \frac{aa P_{II}}{1+\exp(bb/a)} +  F \right] (1-F) $$
that is, the ``invisible'' population is proportional to the seed population that knows the rumor at $t=0$.

\section{Discussion}
\label{Sect_Discussion}

For further clarification, we first summarize the procedure developed above to obtain the function $F(t)$, which depends only on parameters $P_{IP}$, $P_{II}$ and $P_{USG}$. Next, we discuss our model and its capability to make predictions about the phenomenon of spreading rumors.  

We outline below the steps necessary to adjust the experimental data and find the function $F(t)$.

\begin{description}
    \item[First step:] A partition of the parameter space is defined. For each of the 500 points in the partition, numerical simulations are performed and the results obtained are analyzed as indicated below.
    \item[Second step:] Coefficients $a$ and $\epsilon$ are computed using (\ref{eq:fit_num_Fn}). $C$ is estimated using (\ref{eq:fit_num_C}), so that the initial conditions are fulfilled.
    \item[Third step:] We study the dependence of these coefficients on the parameters $P_{IP}$ and $P_{II}$ (for $P_{USG}$ fixed), using (\ref{eq:fit_num_ep}) and (\ref{eq:fit_num_1a}).
    \item[Fourth step:] Using the data set that was fitted in the first steps, we deduce a general expression of the function $F(t)$ and also analyze its dependence on $P_{USG}$ (see (\ref{eq:fit_num_USG})).
\end{description}

In section \ref{sec:USGneq0}, the distribution of burned individuals, $F(t)$, was deduced as a function dependent only on the parameters inherent to the network,  $P_{IP}$, $P_{II}$ and $P_{USG}$. This function becomes a fundamental tool for predicting how long fake news will take to spread and cause problems or for estimating the time necessary for messages, as for instances a marketing campaign or specific information, to be disseminated among the target audience. 

\subsection{Time necessary for a rumor to reach a fraction of the population}

One of the main results of this study is that the function $F(t)$ can be used to estimate the number of burned individuals at any time. An immediate consequence, no less important, is that the time necessary for a rumor to reach a given fraction of the population can also be calculated from this function. In summary, and more formally, the number of iterations $t_X$ (or $n_X$, for discrete time in our model) that are necessary for a rumor to reach a fraction of the population $X$, assuming that the message begins to spread at time $t = 0$, can be calculated using (\ref{eq:fit_num_Fn}).
\begin{equation}
    t_X = F^{-1}( X ) 
    = b+\frac{a}{1+\epsilon} \log\left( \frac{1+\frac{2\epsilon}{1-\epsilon} X}{(1-X) C}\right)
\end{equation}
Therefore, the equation $X=F(t_X)$ provides an implicit relation between the parameters inherent in the network, $P_{II}$, $P_{IP}$ and $P_{USG}$, and the variables that define the necessary time $t_X$ and the fraction of the population $X$. As a result, we can estimate the size of the seed (that is, the value of $P_{II}$) such that a given fraction $X$ of the population is reached at time $t_X$ fixed, in a network characterized by $P_{IP}$ and $P_{USG}$. In particular, the effect of the seed ($P_{II}$) and the size of the uncritical senders group ($P_{USG}$) can be studied to obtain the result sought in a given population.

\subsection{Time evolution from observed data}

The function that models the propagation of a rumor in a network depends, obviously, on the parameters associated with that particular rumor. Our model can be used to estimate these parameters from empirical observations on how the rumor is spread. In fact, only four observations of the number of individuals burned at four different times, since the beginning of the propagation of a rumor, are sufficient to determine the coefficients that fit the function $F (t)$. Then, we can use this adjusted expression to calculate the number of individuals burned at any time in the future.

\begin{figure}
    \centering
    \begin{tabular}{ccc}
    \tiny{$P_{USG}=0\%$} & \tiny{$P_{USG}=3\%$} & \tiny{$P_{USG}=7\%$} \\
    \includegraphics[width=0.31\textwidth]{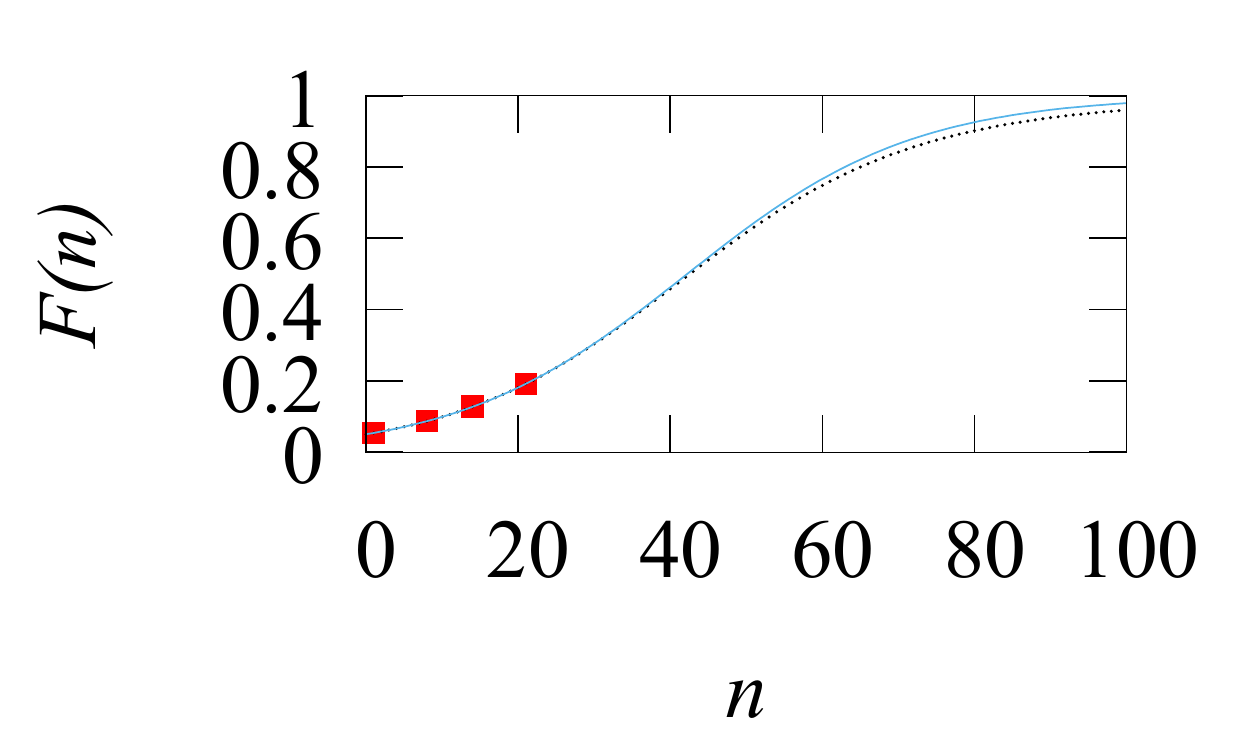} &
    \includegraphics[width=0.31\textwidth]{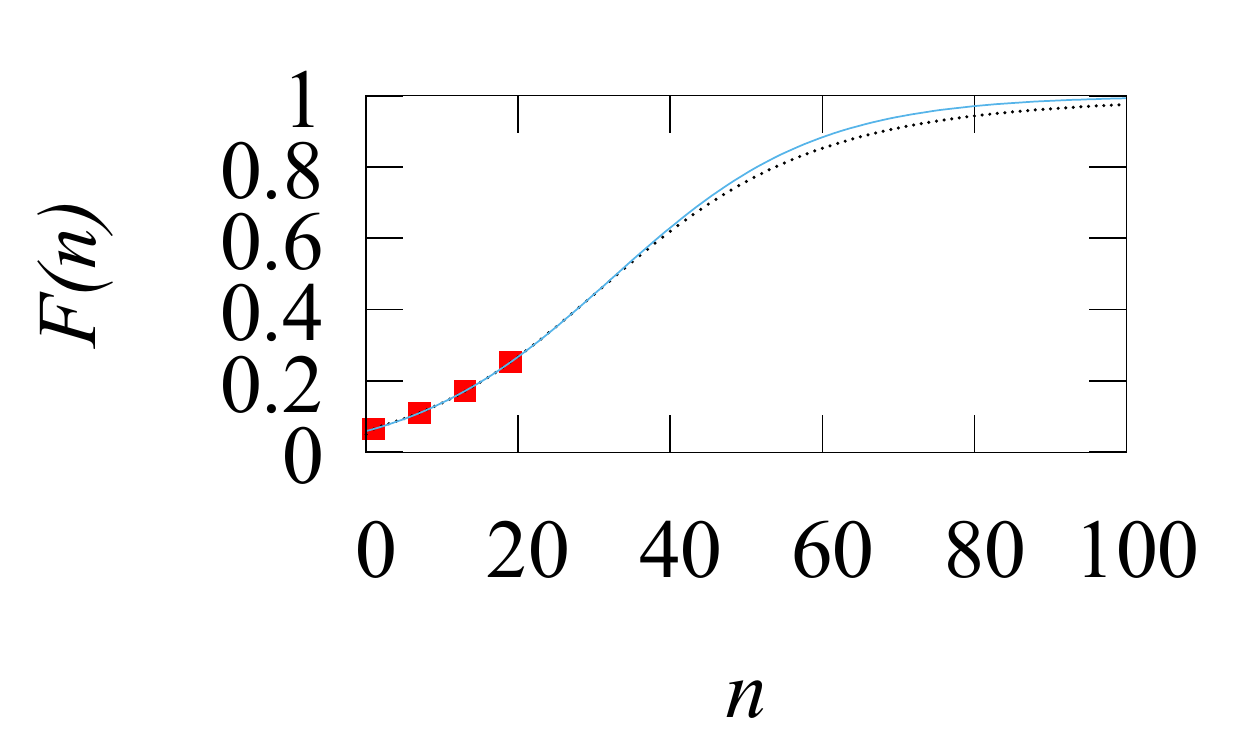} &
    \includegraphics[width=0.31\textwidth]{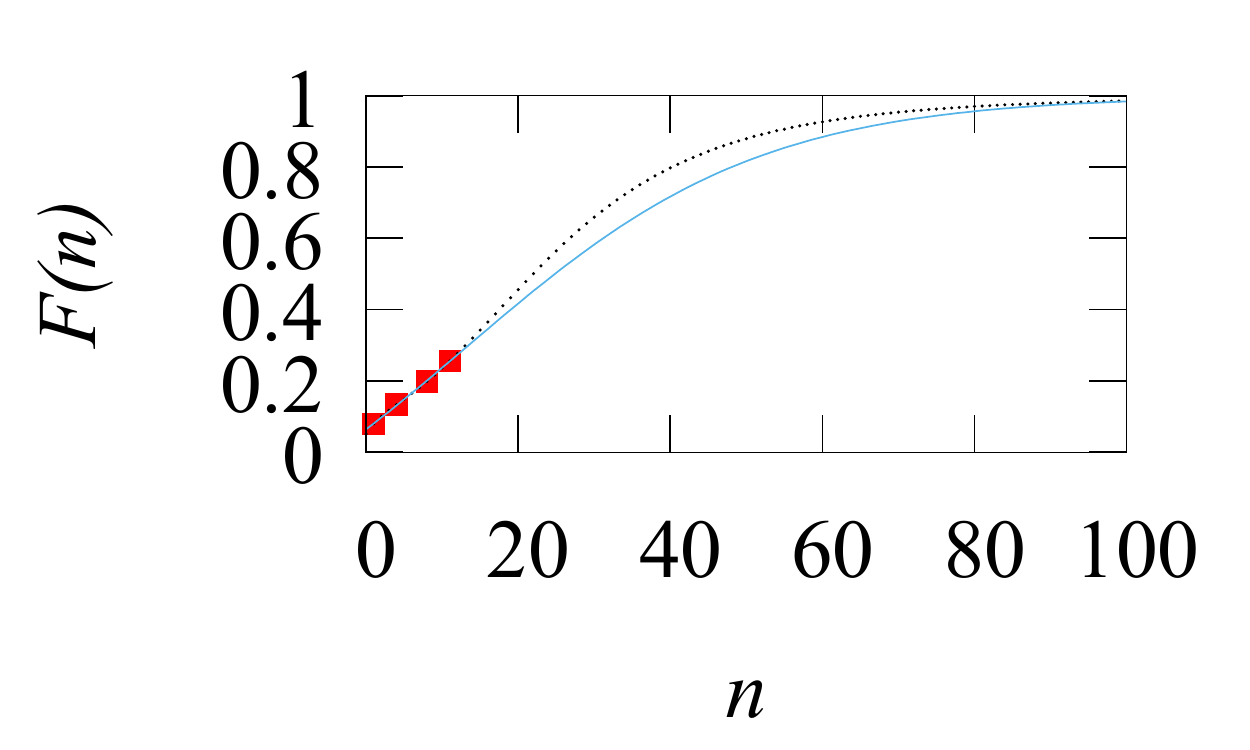}
    \end{tabular}
    \caption{\label{fig:evol_predict} Approximations of $(t,F(t))$ based on the evolution of the early times.  From left to right, the values of $P_{USG}$ are $0\%$, $3\%$ and $7\%$. Fixed parameter: $P_{II}=5\%$ and $P_{IP}=1\%$. The dotted line (green) corresponds to numerical simulation data $(n,f(n))$. The four points used to approximate the coefficients of $F(t)$ are highlighted in red.}
\end{figure}

As it is shown in figure \ref{fig:evol_predict}, the fitted function (blue line) accurately follows that obtained by numerically fitting the data (green dots). These approximations work for different values of $P_{II}$. It should be noted that the adjustment of the coefficients is done for a burned fraction of the population of less than $20\%$, so the function thus fitted can be used to predict when a catastrophe will occur (i.e., a large-scale spreading of an idea). Last but not least, the adjusted coefficients are obtained not only from small values of individuals burned in the population but, and this is the most important, from values that remain almost unchanged over time, see figure \ref{fig:evol_predict}.

\subsection{Network structure inference} \label{sec:Inferencia_parametros}

Knowing the number of burned individuals as a function of spreading time allows us to infer the very structure of the network in which the news is being disseminated. Once the propagation of a rumor for a given topic is known, its current evolution $f(n)$ is used to estimate the coefficients $a$, $\epsilon$ and $C$ in (\ref{eq:fit_num_Fn}), and from them the parameters of the network, $P_{II}$, $P_{IP}$ y $P_{USG}$, can be calculated.

From equations (\ref{eq:fit_num_ep}) and (\ref{eq:fit_num_USG}), the value of the parameter $P_{USG}$ is calculated. Next, when entering in the function $F(t)$ the initial value of $P_{II}$, the estimated value $P_{USG}$ and the number of individuals burned $X$ at time $t_X$, we obtain the value of parameter $P_{IP}$.

Note that the calculation of $P_ {II} $ and $P_ {USG}$ has to be performed for each type of news, because each topic will have its own values for the parameters $P_ {II}$ and $P_{USG}$. Once these parameters are determined, the function $F (t)$ can be used to predict the propagation of similar topic news, and use that prediction to design countermeasures, either to avoid further propagation, e.g., from a political point of view, or to improve the dissemination of news, e.g., with commercial or public information, such as the annual flu vaccination campaign.

\section{Conclusions}
\label{Sect_Conclusions}

As far as we know, we present here a new approach to study the crucial phenomena of rumor propagation through WhatsApp. It is noteworthy that the results obtained in this paper, as well as our algorithms and main techniques used, can be naturally extended to any other type of similar instant messaging application. 

One of our main contributions is to provide a manageable function of time that describes the time evolution of rumor spreading taking into account only person-to-person relationships and contact groups. To be more precise, fixed the initial and propagation probabilities for each individual ($P_{II}$ and $P_{IP}$), the function in (\ref{eq:fit_num_Fn}) defines the fraction of the population who knows the rumor at different times in terms of the parameters that characterize the structure of the network and the dynamics of propagation. In fact, the values of these parameters were estimated from the data obtained by numerical simulations for different scenarios depending on the size of the uncritical senders group ($P_{USG}$).

Conversely, we show that, given the temporal evolution of a rumor by a function as in (\ref{eq:fit_num_Fn}), it can be predicted how the message propagates throughout the network and how much time it takes to burn a fraction of the population. As a set of few parameters ($P_{II}$, $P_{IP}$ and $P_{USG}$) typifies the structure of the network, how a rumor spreads in any social network can be simulated simply by replacing different values of these parameters in the analytical expressions we provide in this paper. Therefore, this information can be used to assess how long it will take a rumor to become dangerous and, consequently, to make decisions accordingly during that time in order to avoid the damages caused by the disinformation. Moreover, the number of individuals that should be burnt to ensure that the rumor reaches a fixed fraction of the population in a given time can be calculated.

Finally, we also study the impact of the so-called uncritical senders group, i.e. a group of individuals who automatically forward a message as soon as they receive it. For all we know, this is a novel idea that allows us to simulate the behavior of groups of highly polarized humans that disseminate fake news with the vile purpose of influencing and causing major changes in society. From our model, we can detect and estimate the size of a group of uncritical senders in a social network by analyzing in particular how the presence of this group changes the temporal evolution in the propagation of a rumor. 

In summary, we present a model that shows how some few but key parameters influence the spread of a rumor and determine the speed with which a rumor may reach a large part of the population. Furthermore, we study how this rumor propagation may be manipulated both through information (or disinformation campaigns) and a group of uncritical senders that actively disseminate some types of news to the entire population.


\begin{thebibliography}{10}

\bibitem{FigOli2017} Figueira A., Oliveira L. The current state of fake news: challenges and opportunities. Procedia Computer Science 121 (2017) 817-825. \\\href{https://doi.org/10.1016/j.procs.2017.11.106}{doi:10.1016/j.procs.2017.11.106} 

\bibitem{MetaxasMustafaraj2010}Metaxas P. \& Mustafaraj E. (2010) From Obscurity to Prominence in Minutes: Political Speech and Real-Time Search. WebSci10: Extending the Frontiers of Society On-Line.

\bibitem{AllGen2017} Allcott H., Gentzkow M. Social media and fake news in the 2016 Election. Journal of Economic Perspectives 31(2) (2017) 211-236. \\\href{https://doi.org/10.1257/jep.31.2.211}{doi:10.1257/jep.31.2.211} 

\bibitem{SilSin2016} Silverman C., Singer-Vine J. Most Americans who see fake news believe it, new survey says. BuzzFeed News (2016). \\\href{https://www.buzzfeednews.com/article/craigsilverman/fake-news-survey}{https://www.buzzfeednews.com/article/craigsilverman/fake-news-survey}

\bibitem{Tambuscioetal2015}  Tambuscio M., Ruffo G., Flammini A., Menczer F. (2015). Fact-checking effect on viral hoaxes: A model of misinformation spread in social networks. In Proceedings of the 24th International Conference on World Wide Web (ACM, 2015), pp. 977–982. \href{http://dx.doi.org/10.1145/2740908.2742572}{doi:10.1145/2740908.2742572} 

\bibitem{Jinetal2013} Jin, Fang and Dougherty, Edward and Saraf, Parang and Cao, Yang and Ramakrishnan, Naren. Epidemiological Modeling of News and Rumors on Twitter. Proceedings of the 7th Workshop on Social Network Mining and Analysis. SNAKDD '13. pp. 8:1--8:9. ACM, New York, 2013. \\\href{https://doi.org/10.1145/2501025.2501027}{doi:10.1145/2501025.2501027}

\bibitem{Kitsatetal2010} Kitsak M, Gallos L.K., Havlin S., Liljeros F., Muchnik L., Stanley H.E. and Makse H.A. (2010). Identification of influential spreaders in complex networks. Nature Physics volume 6, pages 888-893 \\\href{https://www.nature.com/articles/nphys1746}{https://www.nature.com/articles/nphys1746}

\bibitem{BorgeHolthoeferetal2012} Borge-Holthoefer, Meloni S., Gonçalves B., Moreno Y (2012). Emergence of Influential Spreaders in Modified Rumor Models. Journal of Statistical Physics, 151, pages 383-393 \\\href{https://link.springer.com/article/10.1007\%2Fs10955-012-0595-6}{https://link.springer.com/article/10.1007\%2Fs10955-012-0595-6}

\bibitem{DelVicarioetal2016} Del Vicario M., Bessi A., Zollo F., Petroni F., Scala A., Caldarelli G., Stanley H.E., and Quattrociocchi W. (2016). The spreading of misinformation online. PNAS, 113, pages 554-559 \\\href{https://www.pnas.org/content/pnas/113/3/554.full.pdf}{https://www.pnas.org/content/pnas/113/3/554.full.pdf}

\bibitem{Doerretal2012} Doerr B., Fouz M., Friedrich T. (2012). Why rumors spread so quickly in social networks. Communications of the ACM, 55, pages 70-75 \\\href{https://dl.acm.org/citation.cfm?id=2184338}{https://dl.acm.org/citation.cfm?id=2184338}

\bibitem{DaleyKendall1965}Daley D.J., Kendall D.G. (1965). Stochastic Rumours. IMA Journal of Applied Mathematics, 1, pages 42–55 \\\href{https://doi.org/10.1093/imamat/1.1.42}{https://doi.org/10.1093/imamat/1.1.42}

\bibitem{VosRoyAra2017} Vosoughi S., Roy D., Aral S. The spread of true and false news online. Science 359 (2018) 1146-1151. \href{https://doi.org/10.1126/science.aap9559}{doi:10.1126/science.aap9559} 

  
\end{thebibliography}
\end{document}